\documentclass[preprint,showpacs,nofootinbib,aps]{revtex4}
 \usepackage{graphicx}
 \usepackage{bm}

\begin{document}
 \title{Study of the $B_{c}$ ${\to}$ $B_{s}{\pi}$ decay
        with the perturbative QCD approach}
 \author{SUN Junfeng}
 \affiliation{Institute of Particle and Nuclear Physics,
              Henan Normal University, Xinxiang 453007, China}
 \author{YANG Yueling}
 \email{yangyueling@htu.cn}
 \affiliation{Institute of Particle and Nuclear Physics,
              Henan Normal University, Xinxiang 453007, China}
 \author{LU Gongru}
 \affiliation{Institute of Particle and Nuclear Physics,
              Henan Normal University, Xinxiang 453007, China}
 \begin{abstract}
 The $B_{c}$ ${\to}$ $B_{s}{\pi}$ decay is studied with the
 perturbative QCD approach. Three types of wave functions for
 $B_{s}$ meson are considered. The transition form factor
 $F_{0}^{B_{c}{\to}B_{s}}(0)$ and the branching ratio
 ${\cal B}r(B_{c}{\to}B_{s}{\pi})$ are sensitive to the model
 of the $B_{s}$ meson wave functions. With appropriate inputs,
 our estimate on ${\cal B}r(B_{c}{\to}B_{s}{\pi})$ is comparable
 with the recent LHCb measurement. A clear signal of $B_{c}$
 ${\to}$ $B_{s}{\pi}$ decay should be easily observed at the
 Large Hadron Collider.

 \end{abstract}
 \maketitle

 \section{introduction}
 \label{sec1}
 The $B_{c}$ meson, the heaviest of the ground pseudoscalar
 meson with explicit both bottom and charm quantum numbers,
 was observed for the first time via the decay $B_{c}$
 ${\to}$ $J/{\psi}{\ell}{\nu}$ in $1.8$ TeV $p{\bar{p}}$
 collisions using the CDF detector at the Fermilab Tevatron
 in 1998 \cite{CDF.1998}.
 The mass is currently measured at the ${\cal O}(10^{-4})$
 level \cite{prl.100.182002,prl.109.232001}.
 However, there is about $10\%$ uncertainty in the present
 $B_{c}$ lifetime measurement \cite{prl.102.092001,prl.97.012002}.

 The $B_{c}$ meson, laying below $BD$ threshold, is stable for both
 the strong and electromagnetic annihilation interactions.
 It can decay only via the weak interaction.
 The decay modes can be divided into three classes (taking
 $B_{c}^{+}$ as an illustration) \cite{zpc.51.549,prd.49.3399}:
 (1) the $c$ quark decays while the $\bar{b}$ quark as a spectator,
     that is, $c$ ${\to}$ $W^{+}$ $+$ $s$ (or $d$);
 (2) the $\bar{b}$ quark decays while the $c$ quark as a spectator,
     that is, $\bar{b}$ ${\to}$ $W^{+}$ $+$ $\bar{c}$ (or $\bar{u}$);
 (3) the annihilation channel,
     that is, $c$ $+$ $\bar{b}$ ${\to}$ $W^{+}$;
 where the virtual $W^{+}$ boson materializes either into a pair
 of leptons ${\ell}^{+}{\nu}$, or into a pair of quarks which
 then hadronizes.
 Both the $c$ and $\bar{b}$ quarks in $B_{c}^{+}$ meson can
 decay individually, resulting in that the $B_{c}$ lifetime
 is about three times less than other ground pseudoscalar
 $b$-flavored mesons, that is, ${\tau}_{B_{c}}$ $<$
 $\frac{1}{3}{\tau}_{B_{u,d,s}}$ \cite{pdg2012}.
 Enhanced by the hierarchy of the Cabibbo-Kobayashi-Maskawa
 (CKM) quark-mixing matrix elements ${\vert}V_{cs}{\vert}$
 $>$ ${\vert}V_{cb}{\vert}$, the class (1) decay contributes
 ${\sim}$ $70\%$ to the $B_{c}$ decay width \cite{prd.53.4991},
 which shows the critical contribution of the charm quark to
 the $B_{c}$ lifetime.
 In this paper, we study the $B_{c}$ ${\to}$ $B_{s}{\pi}$
 decay which belongs to class (1).
 Our motivations are listed below.

 Firstly, from an experimental consideration.
 It is estimated \cite{pan.67.1559} that with large
 production cross section ${\sigma}(B_{c})$ ${\sim}$
 $1{\mu}{\rm b}$ (taking into account the cascade
 decays of the excited states $B_{c}^{\ast}$ below
 $BD$ threshold) and high luminosity of
 ${\cal L}$ $=$ $10^{34}{\rm cm}^{-2}{\rm s}^{-1}$,
 one could expect $5{\times}10^{10}$ of $B_{c}$ meson
 per year at LHC.
 In addition, LHCb, a dedicated $b$-physics precision
 experiment at LHC, allows proper time resolution of
 ${\sim}$ $50$ fs \cite{1105.5330} which is one order
 of magnitude shorter than lifetime of the $B_{c,s}$ meson.
 This property facilitates the separation between
 primary and decay vertices.
 The rate for the charged pion identification is larger
 than $90\%$ \cite{1105.5330}.
 There seems to be a golden opportunity and a real
 possibility to investigate the $B_{c}$ weak decay.
 The reconstructed events of the $B_{c}$ ${\to}$ $B_{s}{\pi}$
 decay is estimated to be ${\sim}$ $10^{3}$ per year
 \cite{pan.67.1559}.
 Recently, the LHCb collaboration has presented research
 that the $B_{c}$ ${\to}$ $B_{s}{\pi}$ decay is observed with
 significance in excess of five standard deviations based on
 3 fb$^{-1}$ data sample \cite{1308.4544}.

 Secondly, from phenomenological considerations.
 With abundant measurements, we can carefully test
 various theoretical models.
 Accordingly, accurate theoretical prediction
 of $B_{c}$ decay is essential,
 although the $B_{c}$ weak decay is complicated because
 of strong interaction effects.
 Based on an expansion in ${\alpha}_{s}/{\pi}$ and
 ${\Lambda}_{\rm QCD}/m_{Q}$ (where ${\alpha}_{s}$,
 ${\Lambda}_{\rm QCD}$ and $m_{Q}$ are the strong
 coupling constant, QCD characteristic scale and
 mass of heavy quark $Q$, respectively),
 several attractive methods have been proposed to
 evaluate the hadronic matrix elements,
 such as the QCD factorization \cite{qcdf},
 perturbative QCD method (pQCD) \cite{pqcd},
 and soft and collinear effective theory \cite{scet}.
 These approaches have been applied to $B_{u,d}$
 hadronic decays with reasonable explanation for
 measurements.
 Whether or not these approach is suitable for $B_{c}$
 ${\to}$ $B_{s}{\pi}$ decay needs to be examined.
 Herein we investigate $B_{c}$ ${\to}$
 $B_{s}{\pi}$ decay with the pQCD approach to
 give an estimate of the associated branching ratio.

 \section{effective Hamiltonian}
 \label{sec2}
 For hadronic decays, typically the effective Hamiltonian
 calculations with operator product expansion scheme is
 used. The effective Hamiltonian for $B_{c}$ ${\to}$
 $B_{s}{\pi}$ decay can be written as \cite{rmp68p1125}:
  \begin{equation}
 {\cal H}_{\rm eff}=
  \frac{G_{\rm F}}{\sqrt{2}}V_{ud}V_{cs}^{\ast}
  \Big\{C_{1}({\mu})(\bar{s}_{\alpha}c_{\alpha})_{V-A}
             (\bar{u}_{\beta}d_{\beta})_{V-A}
       +C_{2}({\mu})(\bar{s}_{\alpha}c_{\beta})_{V-A}
             (\bar{u}_{\alpha}d_{\beta})_{V-A}\Big\}
  +{\rm H.c.},
  \label{hamiltonian}
  \end{equation}
 where $G_{\rm F}$ is Fermi coupling constant.
 The CKM factor $V_{ud}V_{cs}^{\ast}$ ${\sim}$ ${\cal O}(1)$.
 ${\alpha}$ and ${\beta}$ are color indices.
 $(\bar{q}q^{\prime})_{V-A}$ $=$
 $\bar{q}{\gamma}_{\mu}(1-{\gamma}_{5})q^{\prime}$.
 The Wilson coefficients $C_{1,2}({\mu})$, incorporating the physics
 contributions from heavy particles such as $W$ and $Z$ bosons, top
 quark, and scales higher than ${\mu}$, have been calculated to the
 next-to-leading order with perturbation theory \cite{rmp68p1125}.
 There most difficult problem remaining in calculation is how to
 evaluate the hadronic matrix elements of the local operators
 properly and accurately.

 \section{Hadronic matrix elements}
 \label{sec3}
 Using Brodsky-Lepage approach \cite{prd22p2157},
 the hadronic matrix element is commonly written as
 a convolution of hard-scattering kernels and hadron
 wave functions (WFs).
 It is shown that \cite{qcdf} due to appearance of the
 endpoint divergences, the QCDF formalisms based on the
 collinear approximation fail to give the contributions
 of the spectator and annihilation interactions
 satisfactorily.
 The pQCD approach advocates that \cite{pqcd} the endpoint
 singularity in collinear approximation could be smeared
 by retaining the transverse momentum $k_{T}$ of quarks
 and by introducing the Sudakov factor $e^{-S}$.
 Using the pQCD formalisms, one decay amplitude is
 factorized into three factors: the ``harder'' effects
 incorporated into the Wilson coefficients $C$, the heavy
 quark decay subamplitude $H$, and the universal
 hadron WFs ${\Phi}$, such that:
  \begin{equation}
 {\int}{\bf d}k\,C(t)H(k,t){\Phi}(k)e^{-S}
  \label{am01}
  \end{equation}
 where $k$ and $t$ are the corresponding kinematic variables
 and characteristic scale, respectively.

 \subsection{Kinematic variables}
 \label{sec31}
 In the terms of the light cone coordinate,
 the momenta of the valence quarks and hadrons in the
 rest frame of the $B_{c}$ meson are defined as:
 \begin{eqnarray}
 p_{1}&=&\frac{m_{1}}{\sqrt{2}}(1,1,\vec{0}_{T}),
 \label{momentum-p1} \\
 p_{2}&=&\frac{m_{1}}{\sqrt{2}}(r_{B_{s}}^{2},1,\vec{0}_{T}),
 \label{momentum-p2} \\
 p_{3}&=&\frac{m_{1}}{\sqrt{2}}(1-r_{B_{s}}^{2},0,\vec{0}_{T}),
 \label{momentum-p3} \\
 k_{i}&=&x_{i}p_{i}+(0,0,\vec{k}_{iT}),
 \label{momentum-ki}
 \end{eqnarray}
 where the subscript $i$ $=$ $1$, $2$, $3$ refer to
 $B_{c}$, $B_{s}$, ${\pi}$ meson, respectively.
 Variables $k_{i}$, $\vec{k}_{iT}$, $x_{i}$ are the
 four-dimensional momentum, transverse momentum and
 longitudinal momentum fraction of light quark,
 respectively.
 $r_{B_{s}}$ $=$ $m_{B_{s}}/m_{B_{c}}$.

 \subsection{Wave functions}
 \label{sec32}
 Hadron wave functions are basic input in Eq.(\ref{am01}).
 Adopting the notation in
 \cite{npb592p3,plb523p111,npb625p239,prd55p272,
 prd65p014007,prd74p014027,epjc28p515,jhep9901p010,
 prd.83.094031,jhep0804p061},
 the two-valence-particle WFs for double-light pion,
 heavy-light $B_{s}$ meson and double-heavy $B_{c}$
 meson are decomposed into :
  \begin{eqnarray}
 {\langle}{\pi}(p_{3}){\vert}
  \bar{u}_{\alpha}(z)d_{\beta}(0)
 {\vert}0{\rangle} &=&
  \frac{-i}{\sqrt{2N_{c}}}
 {\int}{\bf d}^{4}k_{3}\,e^{+ik_{3}{\cdot}z}
  \Big\{ {\gamma}_{5}\Big[
  \!\!\not{\!p}_{3}{\phi}_{\pi}^{a}\!
 +\!{\mu}_{\pi}{\phi}_{\pi}^{p}\!
 +\!{\mu}_{\pi}\Big(\!\!\not{\!n}_{+}\!\!\not{\!n}_{-}\!\!
 -\!1\Big){\phi}_{\pi}^{t}
  \Big]\Big\}_{{\beta}{\alpha}},
  \label{wf-pi-01} \\
 {\langle}B_{s}(p_{2}){\vert}
  \bar{s}_{\alpha}(z)b_{\beta}(0)
 {\vert}0{\rangle} &=&
  \frac{-i}{\sqrt{2N_{c}}}
 {\int}{\bf d}^{4}k_{2}\,e^{+ik_{2}{\cdot}z}
  \Big\{ {\gamma}_{5}
  \Big(\!\!\not{\!p}_{2}\!+\!m_{B_{s}}\!\Big)
  \Big[\frac{\not{\!n}_{+}}{\sqrt{2}}{\phi}_{B_{s}}^{-}\!
      +\frac{\not{\!n}_{-}}{\sqrt{2}}{\phi}_{B_{s}}^{+}
  \Big]\Big\}_{{\beta}{\alpha}},
  \label{wf-bs-01} \\
 {\langle}0{\vert}
  \bar{b}_{\alpha}(0)c_{\beta}(z)
 {\vert}B_{c}(p_{1}){\rangle} &=&
  \frac{-i}{\sqrt{2N_{c}}}
 {\int}{\bf d}^{4}k_{1}\,e^{-ik_{1}{\cdot}z}
  \Big\{\Big[
  \frac{\not{\!n}_{+}}{\sqrt{2}}{\phi}_{B_{c}}^{+}\!
  +\frac{\not{\!n}_{-}}{\sqrt{2}}{\phi}_{B_{c}}^{-}
  \Big]\Big(\!\!\not{\!p}_{1}\!+\!m_{B_{c}}\!\Big)
 {\gamma}_{5}\Big\}_{{\beta}{\alpha}},
  \label{wf-bc-01}
  \end{eqnarray}
 where $N_{c}$ $=$ $3$ is the color number.
 $n_{-}$ and $n_{+}$ are the minus and plus null vectors,
 respectively. $n_{+}{\cdot}n_{-}$ $=$ $1$.

 For the heavy-light $B_{s}$ meson, there are two
 scalar WFs ${\phi}^{+}_{B_{s}}$ and ${\phi}^{-}_{B_{s}}$.
 The equation of motion for WFs ${\phi}_{B_{s}}^{\pm}$ is
 \cite{npb592p3,plb523p111}
  \begin{equation}
 {\phi}_{B_{s}}^{+}(x)+x\,{\phi}_{B_{s}}^{-{\prime}}(x)=0
  \label{eom}.
  \end{equation}
 The relationship is helpful in constraining
 models for the $B_{s}$ WFs, which leads to
 ${\phi}_{B_{s}}^{+}(x)$ vanished at the endpoint and
 ${\phi}_{B_{s}}^{-}(x)$ $=$ ${\cal O}(1)$ for
 $x$ ${\to}$ $0$ \cite{npb625p239}.
 Here we will investigate three candidates of WFs
 for $B_{s}$ meson.

 The first candidate is the exponential type (GN) suggested
 in \cite{prd55p272},
  \begin{eqnarray}
 {\phi}_{B_{s}}^{+}(x,b)&=&
  \frac{f_{B_{s}}}{2\sqrt{2N_{c}}}N_{\rm GN}^{+}\,x
 {\exp}\Big[-\frac{x\,m_{B_{s}}}{{\omega}_{\rm GN}}\Big]
  \frac{1}{1+(b\;{\omega}_{\rm GN})^{2}}
  \label{wf-bsp-01}, \\
 {\phi}_{B_{s}}^{-}(x,b)&=&
  \frac{f_{B_{s}}}{2\sqrt{2N_{c}}}N_{\rm GN}^{-}
 {\exp}\Big[-\frac{x\,m_{B_{s}}}{{\omega}_{\rm GN}}\Big]
  \frac{1}{1+(b\;{\omega}_{\rm GN})^{2}}
  \label{wf-bsm-01}.
  \end{eqnarray}

 The second candidate is the Gaussian type (KLS) proposed in
 \cite{prd65p014007,prd74p014027} whereby
  \begin{eqnarray}
 {\phi}_{B_{s}}^{+}(x,b)&=&
  \frac{f_{B_{s}}}{2\sqrt{2N_{c}}}N_{\rm KLS}^{+}x^{2}\bar{x}^{2}\,
 {\exp}\Big[-\frac{1}{2}\Big(\frac{x\,m_{B_{s}}}{{\omega}_{\rm KLS}}\Big)^{2}
            -\frac{1}{2}{\omega}_{\rm KLS}^{2}b^{2}\Big]
  \label{wf-bsp-02}, \\
 {\phi}_{B_{s}}^{-}(x,b)&=&
  \frac{f_{B_{s}}}{2\sqrt{2N_{c}}}N_{\rm KLS}^{-}
 {\exp}\Big[-\frac{1}{2}{\omega}_{\rm KLS}^{2}b^{2}\Big]\Big\{
 {\exp}\Big[-\frac{1}{2}\Big(\frac{x\,m_{B_{s}}}{{\omega}_{\rm KLS}}\Big)^{2}
  \Big]\Big(m_{B_{s}}^{2}\bar{x}^{2}+2{\omega}_{\rm KLS}^{2}\Big)
  \nonumber \\ & &~~~~~~~
 +\sqrt{2{\pi}}m_{B_{s}}{\omega}_{\rm KLS}{\rm Erf}
  \Big(\frac{x\,m_{B_{s}}}{\sqrt{2}{\omega}_{\rm KLS}}\Big)+C_{\rm KLS}\Big\}
  \label{wf-bsm-02},
  \end{eqnarray}
 where $\bar{x}$ $=$ $1$ $-$ $x$, and the constant $C_{\rm KLS}$ is
 chosen so that ${\phi}_{B_{s}}^{-}(1,b)$ $=$ $0$.

 The third candidate is the KKQT type derived form the QCD
 equations of motion and heavy-quark symmetry constraint
 \cite{epjc28p515,plb523p111},
  \begin{eqnarray}
 {\phi}_{B_{s}}^{+}(x,b)&=&
  \frac{f_{B_{s}}}{2\sqrt{2N_{c}}}\frac{2x}{{\omega}_{\rm KKQT}^{2}}
 {\theta}(y)J_{0}\Big(m_{B_{s}}b\sqrt{xy}\Big)
  \label{wf-bsp-03}, \\
 {\phi}_{B_{s}}^{-}(x,b)&=&
  \frac{f_{B_{s}}}{2\sqrt{2N_{c}}}\frac{2y}{{\omega}_{\rm KKQT}^{2}}
 {\theta}(y)J_{0}\Big(m_{B_{s}}b\sqrt{xy}\Big)
  \label{wf-bsm-03},
  \end{eqnarray}
 where $y$ $=$ ${\omega}_{\rm KKQT}$ $-$ $x$.

 In the above equations Eq.(\ref{wf-bsp-01}---\ref{wf-bsm-03}),
 $b$ denotes the conjugate variables of the transverse
 momentum of $s$ quark in $B_{s}$ meson.
 There is only one parameter ${\omega}$ for each kind of
 WFs candidate.
 The normalization constants $N^{\pm}$ is related to the
 decay constant $f_{B_{s}}$ through the relation :
  \begin{equation}
 {\int}_{0}^{1}{\phi}_{B_{s}}^{\pm}(x,0){\bf d}x
 =\frac{f_{B_{s}}}{2\sqrt{2N_{c}}}
  \label{wf-bs-02}.
  \end{equation}

 For the double-light pion, the expression of
 ${\phi}_{\pi}^{a,p,t}$ can be found in
 \cite{prd65p014007,prd74p014027,epjc28p515,jhep9901p010,prd.83.094031}.
 For the double-heavy $B_{c}$ meson, it can be described
 approximatively by nonrelativistic dynamics.
 At tree level and in leading order of the expansion in
 the relative velocities, $b$ and $c$ quarks in the $B_{c}$
 meson just share the total momentum according to their
 masses \cite{jhep0804p061},
  \begin{equation}
 {\phi}_{{B}_{c}}^{\pm}(x)=
  \frac{f_{B_{c}}}{2\sqrt{2N_{c}}}\delta(x-r_{c})
  \label{wf-bc-02},
  \end{equation}
 where $f_{B_{c}}$ is the decay constant, and
 $r_{c}$ $=$ $m_{c}/m_{B_{c}}$.

 \section{$B_{c}$ ${\to}$ $B_{s}$ transition form factors}
 \label{sec4}
 The $B_{c}$ ${\to}$ $B_{s}$ transition form factors are
 defined as follows \cite{zpc29p637}:
 \begin{eqnarray}
 & & {\langle}B_{s}(p_{2}){\vert}\bar{s}
     {\gamma}^{\mu}(1-{\gamma}_{5})c
     {\vert}B_{c}(p_{1}){\rangle}
      \nonumber \\
 &=&  \Big[(p_{1}+p_{2})^{\mu}
     -\frac{m_{B_{c}}^{2}-m_{B_{s}}^{2}}{q^{2}}q^{\mu}
      \Big] F_{1}^{B_{c}{\to}B_{s}}(q^{2})
      \nonumber \\
 &+&  \frac{m_{B_{c}}^{2}-m_{B_{s}}^{2}}{q^{2}}q^{\mu}
      F_{0}^{B_{c}{\to}B_{s}}(q^{2})
 \label{ff-01},
 \end{eqnarray}
 where $q$ $=$ $p_{1}$ $-$ $p_{2}$ is the momentum
 transfer.
 Usually, the longitudinal form factor $F_{0}(q^{2})$
 is compulsorily equal to the transverse form factor
 $F_{1}(q^{2})$ in the largest recoil limit to cancel
 singularities appearing at the pole $q^{2}$ $=$ $0$,
 i.e., $F_{0}(0)$ $=$ $F_{1}(0)$.

 Within the pQCD framework, the one-gluon-exchange
 diagrams contributing to the $B_{c}$ ${\to}$ $B_{s}$
 transition form factors are displayed in Fig.\ref{fig01}.
 It has been shown that the pQCD approach works ideally
 in the large recoil region \cite{pqcd,epjc28p515,prd74p014027,npb625p239}.
 The expression of form factors is :
  \begin{eqnarray} &&
    F_{0}^{B_{c}{\to}B_{s}}(0)\,=\,
    F_{1}^{B_{c}{\to}B_{s}}(0)
  \nonumber \\ &=&  8{\pi}m_{B_{c}}^{2}C_{F}
 {\int}_{0}^{1}{\bf d}x_{1}
 {\int}_{0}^{1}{\bf d}x_{2}
 {\int}_{0}^{\infty}\!b_{2}{\bf d}b_{2}
  \nonumber \\ &{\times}&
  \Big\{E_{a}(t_{a}){\alpha}_{s}(t_{a})
        H_{a}({\alpha},{\beta}_{a},b_{2})\,
        r_{B_{s}}{\phi}_{B_{s}}^{+}(x_{2})
  \nonumber \\ &&{\times}
  \Big[{\phi}_{B_{c}}^{+}(x_{1})
        \Big(x_{2}r_{B_{s}}+r_{c}-x_{2}\Big)
  \nonumber \\ &&~
      +{\phi}_{B_{c}}^{-}(x_{1})
        \Big(x_{2}r_{B_{s}}+r_{c}-r_{B_{s}}r_{c}\Big)\Big]
  \nonumber \\ &&
      + E_{b}(t_{b}){\alpha}_{s}(t_{b})
        H_{b}({\alpha},{\beta}_{b},b_{2})\,
        x_{1}{\phi}_{B_{c}}^{-}(x_{1})
  \nonumber \\ && {\times}
  \Big[{\phi}_{B_{s}}^{-}(x_{2})\Big(1-r_{B_{s}}\Big)
      +{\phi}_{B_{s}}^{+}(x_{2})r_{B_{s}}^{2}\Big]\Big\}
  \label{ff-11},
  \end{eqnarray}
 where $C_{F}$ $=$ $4/3$ is color factor.
 $E_{i}(t_{i})$ $=$ $e^{-S_{B_{s}}(t_{i})}$ is the
 Sudakov factor.
 \begin{equation}
  S_{B_{s}}(t)=s(x_{2}p_{2}^{-},b_{2})
  +2{\int}_{1/b_{2}}^{t}\frac{{\bf d}{\mu}}{\mu}{\gamma}_{q}
 \label{ff-13},
 \end{equation}
 where ${\gamma}_{q}$ $=$ $-{\alpha}_{s}/{\pi}$ is the
 quark anomalous dimension.
 The expression of $s(Q,b)$ can be found in
 \cite{prd52p3958,npb642p263}.
 The hard-scattering kernel function $H_{a,b}$
 in Eq.(\ref{ff-11}) is defined as :
 \begin{eqnarray}
 H_{a}({\alpha},{\beta}_{a},b_{2})
 &=&
 \frac{K_{0}(\sqrt{{\alpha}}b_{2})-K_{0}(\sqrt{{\beta}} b_{2})}
      {{\beta}_{a}-{\alpha}}
 \label{ff-14}, \\
 H_{b}({\alpha},{\beta}_{b},b_{2})
 &=&
 \frac{K_{0}(\sqrt{{\alpha}}b_{2})}{{\beta}_{b}}
 \label{ff-15},
 \end{eqnarray}
 where $K_{0}$ and $I_{0}$ are the modified Bessel functions;
 ${\alpha}$ and ${\beta}$ are the virtualities of internal
 gluons and quarks, respectively.
  \begin{eqnarray}
 {\alpha}&=&-m_{B_{c}}^{2}\Big[\bar{x}_{1}^{2}
            +r_{B_{s}}^{2}\bar{x}_{2}^{2}
            -\bar{x}_{1}\bar{x}_{2}\Big(1+r_{B_{s}}^{2}\Big)\Big]
  \label{ff-21}, \\
 {\beta}_{a}&=&-m_{B_{c}}^{2}\Big[1+r_{B_{s}}^{2}\bar{x}_{2}^{2}
               -\bar{x}_{2}\Big(1+r_{B_{s}}^{2}\Big)-r_{c}^{2}\Big]
  \label{ff-22}, \\
 {\beta}_{b}&=&-m_{B_{c}}^{2}\Big[r_{B_{s}}^{2}+\bar{x}_{1}^{2}
               -\bar{x}_{1}\Big(1+r_{B_{s}}^{2}\Big)\Big]
  \label{ff-23}, \\
 t_{i}&=&{\max}(\sqrt{{\vert}{\alpha}{\vert}},
                \sqrt{{\vert}{\beta}_{i}{\vert}},1/b_{2})
  \label{ff-24}.
  \end{eqnarray}

 \section{decay amplitude}
 \label{sec5}
 Within the pQCD framework, the Feynman diagrams for $B_{c}$
 ${\to}$ $B_{s}{\pi}$ decay are shown in Fig.\ref{fig02},
 where (a) and (b) are factorizable topology,
 (c) and (d) are nonfactorizable topology.
 After a straightforward calculation, the decay amplitudes
 is written as :
  \begin{equation}
 {\cal A}(B_{c}{\to}B_{s}{\pi})=
  \frac{G_{F}}{\sqrt{2}}V_{ud}V_{cs}^{\ast}
  \sum\limits_{i=a,b,c,d}{\cal A}_{i}
  \label{am11},
  \end{equation}
 where the explicit expressions of ${\cal A}_{i}$
 are collected in APPENDIX \ref{app01}.
 From these expressions, we see that only the twist-2
 light-cone distribution amplitude (LCDA) ${\phi}_{\pi}^{a}$
 of pion contributes to nonfactorizable decay amplitude.
 The expression of the twist-2 pion LCDA is defined in Gegenbauer
 polynomials \cite{jhep9901p010} as:
  \begin{equation}
 {\phi}_{\pi}^{a}(x,{\mu})=
  \frac{f_{\pi}}{2\sqrt{2N_{c}}}6x\bar{x}
  \Big\{1+a_{2}^{\pi}({\mu})C_{2}^{3/2}(x-\bar{x})
 +a_{4}^{\pi}({\mu})C_{4}^{3/2}(x-\bar{x})\Big\},
  \end{equation}
 where the nonperturbative parameter $a_{i}^{\pi}({\mu})$
 is the Gegenbauer moment.

 The branching ratio in the $B_{c}$ meson rest frame can be written as:
 \begin{equation}
  {\cal B}r(B_{c}{\to}B_{s}{\pi})= \frac{{\tau}_{B_{c}}}{8{\pi}}
   \frac{p}{m_{B_{c}}^{2}}
  {\vert}{\cal A}(B_{c}{\to}B_{s}{\pi}){\vert}^{2}
   \label{eq:br-01},
 \end{equation}
 where the common momentum
 $p$ $=$ $(m_{B_{c}}^{2}-m_{B_{s}}^{2})/2m_{B_{c}}$.

 \section{Numerical results and discussion}
 \label{sec6}
 The input parameters in our numerical calculation are
 collected in Table \ref{tab01}.
 If not specified explicitly, we shall take their
 central values as the default input.

 Our results on form factor $F_{0}^{B_{c}{\to}B_{s}}$(0) and
 branching ratio ${\cal B}r(B_{c}{\to}B_{s}{\pi})$ are
 listed in Table \ref{tab02},
 where the first uncertainty comes from the mass of charm
 quark; the second uncertainty comes from the shape parameter
 of $B_{s}$ meson WFs, that is,
 ${\omega}_{\rm GN}$ $=$ $0.65{\pm}0.10$ GeV in
 Eq.(\ref{wf-bsp-01}-\ref{wf-bsm-01}),
 ${\omega}_{\rm KLS}$ $=$ $0.75{\pm}0.10$ GeV in
 Eq.(\ref{wf-bsp-02}-\ref{wf-bsm-02}) and
 ${\omega}_{\rm KKQT}$ $=$ $0.35{\pm}0.10$ in
 Eq.(\ref{wf-bsp-03}-\ref{wf-bsm-03});
 the third uncertainty comes from the choice of hard
 scales $(1{\pm}0.1)t_{i}$ in Eq.(\ref{ff-24}).
 We can see that the large uncertainty come from
 the $B_{s}$ WFs.
 In addition, the decay constants $f_{B_{c}}$ and
 $f_{B_{s}}$ will bring some $2.4\%$ uncertainty
 to the form factor.

\begin{figure}[ht]
 \includegraphics[angle=0,width=0.70\textwidth,bb=180 675 450 725]{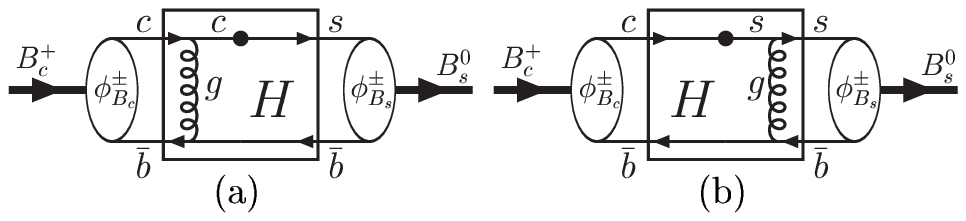}
 \caption{Diagrams contributing to the $B_{c}$ ${\to}$ $B_{s}$ transition
          form factor, where the box represents the lowest order
          hard-scattering kernel $H$, and the dot denotes an appropriate
          Dirac matrix.}
 \label{fig01}
 \end{figure}

 For the form factor $F_{0}^{B_{c}{\to}B_{s}}(0)$,
 our study show that :
  Firstly,
  the interference between Fig.\ref{fig01}(a) and (b)
  is constructive. Compared with Fig.\ref{fig01}(b),
  the contribution of Fig.\ref{fig01}(a) is
  ${\lesssim}$ $30\%$;
  Secondly,
  both ${\phi}_{B_{c}}^{+}$ and ${\phi}_{B_{c}}^{-}$
  contribute to the form factor, and their interference
  is constructive. Compared with ${\phi}_{B_{c}}^{-}$,
  the contribution of ${\phi}_{B_{c}}^{+}$ is
  ${\lesssim}$ $20\%$;
  Thirdly,
  the interference between ${\phi}_{B_{s}}^{+}$ and
  ${\phi}_{B_{s}}^{-}$ is constructive. Compared with
  ${\phi}_{B_{s}}^{+}$, the contribution of ${\phi}_{B_{s}}^{-}$
  is $<$ $20\%$ for both GN and KLS type;
  Fourthly,
  the form factor $F_{0}^{B_{c}{\to}B_{s}}(0)$ is sensitive
  to the shape parameter of $B_{s}$ WFs;
  Fifthly,
  By keeping the parton transverse momentum $k_{T}$,
  and employing the Sudakov factors to suppress the long
  distance contribution in large $b$ region \cite{pqcd},
  the form factor $F_{0}^{B_{c}{\to}B_{s}}(0)$ is
  perturbatively calculable with the pQCD approach.
  The contribution to form factor comes completely from
  ${\alpha}_{s}/{\pi}$ $<$ $0.3$ region with the
  scale of Eq.(\ref{ff-24});
  Lastly,
  considering the uncertainties, our results is reasonable
  agreement with the previous results listed in Table \ref{tab03}.

 For the branching ratio ${\cal B}r(B_{c}{\to}B_{s}{\pi})$,
 our study show that :
 Firstly,
 the dominated contribution comes from the factorizable
 topology Fig.\ref{fig02}(a,b).
 The interference between nonfactorizable diagrams
 Fig.\ref{fig02}(c) and (d) is destructive;
 Secondly,
 the main uncertainty is from $B_{s}$ WFs.
 Considering the uncertainties, our results is basically
 comparable with the previous results listed in
 Table \ref{tab03},
 and is also agreement with recent LHCb estimate
 ${\cal B}r(B_{c}{\to}B_{s}{\pi})$ ${\sim}$ 10\%
 \cite{1308.4544}.

 \section{Summary}
 \label{sec7}
 Herein we consider three models (exponential, Gaussian,
 and KKQT type) of $B_{s}$ meson WFs, and study the $B_{c}$
 ${\to}$ $B_{s}$ transition form factor and the branching ratio
 for the $B_{c}$ ${\to}$ $B_{s}{\pi}$ decay with the pQCD approach.
 We find that
 (1) By keeping the parton transverse momentum $k_{T}$ and employing
 the Sudakov factors, $F_{0}^{B_{c}{\to}B_{s}}(0)$ and
 ${\cal B}r(B_{c}{\to}B_{s}{\pi})$ are perturbatively calculable
 within the pQCD framework. The contribution to form factor comes
 completely from ${\alpha}_{s}/{\pi}$ $<$ $0.3$ region.
 (2) $F_{0}^{B_{c}{\to}B_{s}}(0)$ and ${\cal B}r(B_{c}{\to}B_{s}{\pi})$
 are sensitive to the model and the shape parameter ${\omega}$ of
 the $B_{s}$ meson wave functions.
 (3) With appropriate inputs, our estimates on $F_{0}^{B_{c}{\to}B_{s}}(0)$
 and ${\cal B}r(B_{c}{\to}B_{s}{\pi})$ are comparable with the
 previous results. The branching ratio for the $B_{c}$ ${\to}$ $B_{s}{\pi}$
 decay is about a few per cent, which is agreement with the recent
 LHCb measurement.

 With the running of LHC, the more data will be
 accumulated at LHCb, the more precise branching ratio for the
 $B_{c}$ ${\to}$ $B_{s}{\pi}$ decay will be obtained, and various
 models for hadron wave function and factorization treatments for
 hadron matrix element will be more stringently tested.

 \section*{Acknowledgments}
 This work was supported by {\em National Natural Science Foundation
 of China} under Grant Nos. 11147008, U1232101 and 11275057). We
 thank Prof. Hsiangnan Li, Prof. Caidian l\"{u}, Prof. Zhenjun
 Xiao for their helpful discussion.

 \begin{appendix}
 \section{the decay amplitudes}
 \label{app01}
\begin{figure}[ht]
 \includegraphics[angle=0,width=0.95\textwidth,bb=100 675 500 775]{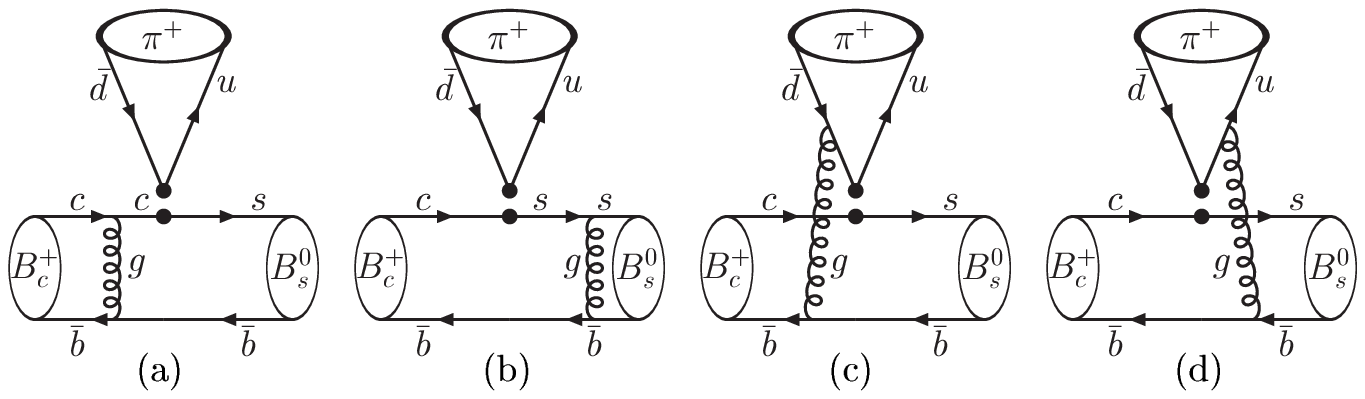}
 \caption{Feynman diagrams for $B_{c}$ ${\to}$ $B_{s}{\pi}$ decay
          within the pQCD framework, where the dots deonte an appropriate
          Dirac matrix.}
 \label{fig02}
 \end{figure}

 There are four diagrams contributing to the $B_{c}$ ${\to}$
 $B_{s}{\pi}$ decay which are shown in Fig.\ref{fig02}.
 The expression of the decay amplitudes in Eq.(\ref{am11}) are :
  \begin{eqnarray}
 {\cal A}_{a} &=&
 -i8{\pi}C_{F}f_{\pi}m_{B_{c}}^{4}
  \Big(1-r_{B_{s}}^{2}\Big)r_{B_{s}}
 {\int}_{0}^{1}{\bf d}x_{1}{\bf d}x_{2}
 {\int}_{0}^{\infty}b_{2}{\bf d}b_{2}
  \nonumber \\ &{\times}&
 E_{a}(t_{a}){\alpha}_{s}(t_{a})
 H_{a}({\alpha},{\beta}_{a},b_{2})
  \Big[C_{1}(t_{a})+\frac{1}{N_{c}}C_{2}(t_{a})\Big]
 {\phi}_{B_{s}}^{+}(x_{2})
  \nonumber \\ &{\times}&
  \Big\{ {\phi}_{B_{c}}^{+}(x_{1})
  \Big[x_{2}r_{B_{s}}+r_{c}-x_{2}\Big]
        +{\phi}_{B_{c}}^{-}(x_{1})
  \Big[x_{2}r_{B_{s}}+r_{c}-r_{B_{s}}r_{c}
  \Big] \Big\}
  \label{amp-11}, \\
 {\cal A}_{b} &=&
 -i8{\pi}C_{F}f_{\pi}m_{B_{c}}^{4}
  \Big(1-r_{B_{s}}^{2}\Big)
 {\int}_{0}^{1}{\bf d}x_{1}{\bf d}x_{2}
 {\int}_{0}^{\infty}b_{2}{\bf d}b_{2}
  \nonumber \\ &{\times}&
 E_{b}(t_{b}){\alpha}_{s}(t_{b})
 H_{b}({\alpha},{\beta}_{b},b_{2})
  \Big[C_{1}(t_{b})+\frac{1}{N_{c}}C_{2}(t_{b})\Big]
  x_{1}{\phi}_{B_{c}}^{-}(x_{1})
 \nonumber \\ &{\times}&
 \Big\{ {\phi}_{B_{s}}^{-}(x_{2})\Big[1-r_{B_{s}}\Big]
       +{\phi}_{B_{s}}^{+}(x_{2})r_{B_{s}}^{2} \Big\}
 \label{amp-12}, \\
 {\cal A}_{c} &=&
  \frac{-i32{\pi}C_{F}m_{B_{c}}^{4}}{\sqrt{2N_{c}}}
  \Big(1-r_{B_{s}}^{2}\Big)r_{B_{s}}
 {\int}_{0}^{1}{\bf d}x_{1}{\bf d}x_{2}{\bf d}x_{3}
 {\int}_{0}^{\infty}b_{2}{\bf d}b_{2}b_{3}{\bf d}b_{3}
  \nonumber \\ &{\times}&
 E_{c}(t_{c}){\alpha}_{s}(t_{c})
 H_{c}({\alpha},{\beta}_{c},b_{2},b_{3})
 C_{2}(t_{c}){\phi}_{B_{s}}^{+}(x_{2})
 {\phi}_{\pi}^{a}(x_{3})
  \nonumber \\ &{\times}&
  \Big\{ {\phi}_{B_{c}}^{+}(x_{1})
          \Big(1-r_{B_{s}}\Big)
          \Big(x_{2}-x_{1}\Big)
        +{\phi}_{B_{c}}^{-}(x_{1})
          \Big[r_{B_{s}}^{2}\Big(x_{2}-x_{3}\Big)
        + \Big(x_{3}-x_{1}\Big) \Big] \Big\}
  \label{amp-13}, \\
 {\cal A}_{d} &=&
  \frac{-i32{\pi}C_{F}m_{B_{c}}^{4}}{\sqrt{2N_{c}}}
  \Big(1-r_{B_{s}}^{2}\Big)
 {\int}_{0}^{1}{\bf d}x_{1}{\bf d}x_{2}{\bf d}x_{3}
 {\int}_{0}^{\infty}b_{2}{\bf d}b_{2}b_{3}{\bf d}b_{3}
  \nonumber \\ &{\times}&
 E_{d}(t_{d}){\alpha}_{s}(t_{d})
 H_{d}({\alpha},{\beta}_{d},b_{2},b_{3})
 C_{2}(t_{d}){\phi}_{B_{c}}^{-}(x_{1})
 {\phi}_{\pi}^{a}(x_{3})
  \nonumber \\ &{\times}&
  \Big\{ {\phi}_{B_{s}}^{+}(x_{2})r_{B_{s}}
          \Big[\Big(x_{1}-\bar{x}_{3}\Big)
 -r_{B_{s}}^{2}\Big(x_{2}-\bar{x}_{3}\Big)\Big]
 -{\phi}_{B_{s}}^{-}(x_{2})
  \Big(1-r_{B_{s}}\Big)\Big(x_{1}-x_{2}\Big)\Big\}
  \label{amp-14},
  \end{eqnarray}
 where the evolution factor $E_{a,b}$ and the kernel function $H_{a,b}$
 are the same as those for the $B_{c}$ ${\to}$ $B_{s}$ transition form
 factors given in Section \ref{sec4}.
 The evolution factor $E_{c,d}$ $=$ $e^{-S_{B_{s}}-S_{\pi}}$.
 The Sudakov factors $S_{B_{s}}$ are given in Eq.(\ref{ff-13}).
 And the Sudakov factors $S_{\pi}$ is defined as :
 \begin{equation}
 S_{\pi}(t)=s(x_{3}p_{3}^{+},b_{3})+s(\bar{x}_{3}p_{3}^{+},b_{3})
           +2{\int}_{1/b_{3}}^{t}\frac{{\bf d}{\mu}}{\mu}{\gamma}_{q}
 \label{am-15}.
 \end{equation}
 The hard-scattering kernel function $H_{c,d}$ is defined as :
  \begin{equation}
  H_{i}({\alpha},{\beta}_{i},b_{2},b_{3})=
  K_{0}(\sqrt{{\beta}_{i}}b_{3})
  \Big\{ {\theta}\Big(b_{2}-b_{3}\Big)
  K_{0}(b_{2}\sqrt{\alpha})
  I_{0}(b_{3}\sqrt{\alpha})
 +b_{2}{\leftrightarrow}b_{3}\Big\}
  \label{am-21},
  \end{equation}
 where ${\alpha}$ and ${\beta}_{i}$ are the virtualities of internal
 gluons and quarks in Fig.\ref{fig02}, respectively;
 $t_{i}$ is the maximal (either longitudinal or transverse) virtuality
 of propagators:
  \begin{eqnarray}
 {\beta}_{c}&=&-m_{1}^{2}\Big(x_{1}-x_{2}\Big)
  \Big[x_{1}-x_{3}-r_{B_{s}}^{2}\Big(x_{2}-x_{3}\Big)\Big]
  \label{am-22}, \\
 {\beta}_{d}&=&-m_{1}^{2}\Big(x_{1}-x_{2}\Big)
  \Big[x_{1}-\bar{x}_{3}-r_{B_{s}}^{2}\Big(x_{2}-\bar{x}_{3}\Big)\Big]
  \label{am-23}, \\
  t_{i}&=&{\max}(\sqrt{{\vert}{\alpha}{\vert}},
  \sqrt{{\vert}{\beta}_{i}{\vert}},1/b_{2},1/b_{3})
  ~~~{\rm for}~~~i=c,d
  \label{am-24}.
  \end{eqnarray}
  \end{appendix}

  \begin{table}[h]
  \caption{input parameters for $B_{c}$  ${\to}$ $B_{s}{\pi}$ decay}
  \label{tab01}
  \begin{ruledtabular}
  \begin{tabular}{l|cc}
  \multicolumn{1}{c|}{parameter} & numerical value & Ref. \\ \hline
  mass of $B_{c}$     & $m_{B_{c}}$ $=$ $6.277{\pm}0.006$~GeV & \cite{pdg2012} \\
  mass of $B_{s}$     & $m_{B_{s}}$ $=$ $5366.7{\pm}0.4$~MeV  & \cite{pdg2012} \\
  mass of $c$ quark   & $m_{c}$ $=$ $1.275{\pm}0.025$~GeV & \cite{pdg2012} \\
  lifetime of $B_{c}$ & ${\tau}_{B_{c}}$ $=$ $0.453{\pm}0.041$~ps  & \cite{pdg2012} \\
  decay constant of ${\pi}$ & $f_{\pi}$ $=$ $130.41{\pm}0.03{\pm}0.20$~MeV & \cite{pdg2012} \\
  decay constant of $B_{c}$ & $f_{B_{c}}$ $=$ $489{\pm}4{\pm}3$~MeV  & \cite{plb651p171} \\
  decay constant of $B_{s}$ & $f_{B_{s}}$ $=$ $227.6{\pm}5.0$~MeV    & \cite{lattice}  \\
  Gegenbauer moment & $a^{\pi}_{2}(1{\rm GeV})=0.16{\pm}0.01$ & \cite{ijmpa25p513} \\
                    & $a^{\pi}_{4}(1{\rm GeV})=0.04{\pm}0.01$ & \cite{ijmpa25p513}
  \end{tabular}
  \end{ruledtabular}
  \end{table}
  \begin{table}[h]
  \caption{Form factor $F_{0}^{B_{c}{\to}B_{s}}(0)$ and
  branching ratio ${\cal B}r(B_c{\to}B_s{\pi})$.}
  \label{tab02}
  \begin{ruledtabular}
  \begin{tabular}{cccc}
 & GN & KLS & KKQT \\ \hline
 $F_{0}$
 & $0.84^{+0.06+0.28+0.09}_{-0.05-0.20-0.05}$
 & $0.87^{+0.08+0.36+0.09}_{-0.07-0.25-0.05}$
 & $0.62^{+0.06+1.08+0.06}_{-0.05-0.37-0.04}$
  \\
  ${\cal B}r{\times}10^{2}$
 & $ 6.20^{+ 0.90+ 4.78+ 1.83}_{- 0.75- 2.60- 0.90}$
 & $ 6.57^{+ 1.19+ 6.57+ 1.90}_{- 0.98- 3.26- 0.94}$
 & $ 3.40^{+ 0.65+22.32+ 0.79}_{- 0.52-~2.84- 0.48}$
  \end{tabular}
  \end{ruledtabular}
  \end{table}

 \begin{table}[htb]
 \caption{Form factor $F_{0}^{B_{c}{\to}B_{s}}(0)$ and
          branching ratio ${\cal B}r(B_{c}{\to}B_{s}{\pi})$
          (in the unit of \%) in the previous literature.}
 \label{tab03}
 \begin{ruledtabular}
 \begin{tabular}{cc|ccc|cc}
    Ref. & $F_{0}$
  & Ref. & $F_{0}$ & ${\cal B}r{\times}10^2$
  & Ref. & ${\cal B}r{\times}10^2$ \\ \hline
    \cite{Phys.Rev.D82.094014.2010}
  & 0.926\footnotemark[1]
  & \cite{Phys.Rev.D86.094028.2012}
  & 1.03
  & 12.01 [10.9]\footnotemark[2]
  & \cite{Phys.Rev.D77.114004.2008}
  & 5.31\footnotemark[3] \\
    \cite{Phys.Rev.D80.074005.2009}
  & 1.021\footnotemark[4]
  & \cite{Phys.Rev.D80.114003.2009}
  & 0.573 [0.571]\footnotemark[5]
  & 3.723 [3.697]\footnotemark[5]
  & \cite{Phys.Rev.D73.054024.2006}
  & 3.9\footnotemark[6] \\
    \cite{Phys.Rev.D79.054012.2009}
  & $0.73^{+0.03+0.03}_{-0.04-0.03}$\footnotemark[7]
  & \cite{Phys.Rev.D79.034004.2009}
  & $0.55{\pm}0.03$\footnotemark[8]
  & $4.8{\pm}0.5$\footnotemark[8]
  & \cite{Phys.Rev.D62.014019.2000}
  & 1.57\footnotemark[9] \\
    \cite{Int.J.Mod.Phys.A23.3237.2008}
  & 1.02\footnotemark[10]
  & \cite{Phys.Rev.D74.074008.2006}
  & $0.58^{+0.01}_{-0.02}$\footnotemark[11]
  & $3.51^{+0.19}_{-0.06}$\footnotemark[11]
  & \cite{Phys.Rev.D56.4133.1997}
  & 4.30\footnotemark[9] \\
    \cite{Mod.Phys.Lett.A16.1439.2001}
  & 0.297\footnotemark[1]
  & \cite{Eur.Phys.J.C32.29.2003}
  & 0.50\footnotemark[6]
  & 2.52\footnotemark[6]
  & \cite{9803433}
  & 5.5\footnotemark[12] \\
    \cite{Phys.Rev.D63.074010.2001}
  & $0.61$\footnotemark[6]
  & \cite{Phys.Atom.Nucl.67.1559.2004}
  & 1.3\footnotemark[13]
  & 16.4\footnotemark[13]
  & \cite{prd.49.3399}
  & 5.93 \\
    \cite{9810339}
  & 0.403${\sim}$0.617\footnotemark[14]
  & \cite{Nucl.Phys.B585.353.2000}
  & 1.3\footnotemark[13]
  & 17.5\footnotemark[13]
  & \cite{zpc.51.549}
  & 2.19\footnotemark[8] \\
    \cite{Z.Phys.C64.57.1994}
  & $0.60{\pm}0.12$\footnotemark[15]
  & \cite{Phys.Rev.D61.034012.2000}
  & 0.66\footnotemark[16]
  & 4.0\footnotemark[16]
  & \cite{zpc.51.549}
  & 3.10\footnotemark[1] \\
    \cite{Z.Phys.C57.43.1993}
  & $0.30{\pm}0.05$\footnotemark[13]
  & \cite{Phys.Usp.38.1.1995}
  & 0.61
  & 3.28\footnotemark[8]
  & \cite{Phys.Usp.38.1.1995}
  & 4.66\footnotemark[1] \\
    \cite{J.Phys.G26.1079.2000}
  & 0.5917\footnotemark[6]
  & \cite{Phys.Rev.D39.1342.1989}
  & 0.340${\sim}$0.925\footnotemark[8]
  & 1.23${\sim}$9.09\footnotemark[8]
  & \\
  &
  & \cite{Phys.Atom.Nucl.62.1739.1999}
  & 0.564\footnotemark[1]
  & 4.37\footnotemark[1]
 \end{tabular}
 \end{ruledtabular}
 {\renewcommand{\baselinestretch}{1}
  \renewcommand\arraystretch{1}
 \footnotetext[1]{It is estimated with the Isgur-Scora-Grinstein-Wise
       \cite{Phys.Rev.D39.799.1989} quark model.}
 \footnotetext[2]{It is estimated with the relativistic quark model,
       where $a_{1}$ $=$ 1.26 [1.2].}
 \footnotetext[3]{It is estimated with the QCD factorization approach
       \cite{qcdf} at the leading order.}
 \footnotetext[4]{It is estimated with a relativistic quark model based
       on a confining potential in the equally mixed scalar-vector
       harmonic form.}
 \footnotetext[5]{It is estimated with the light-front quark model, where
       the interaction is Coulomb plus linear [harmonic oscillator]
       confining potentials.}
 \footnotetext[6]{It is estimated with a relativistic quark model.}
 \footnotetext[7]{It is estimated with covariant light-front quark model,
       where the uncertainties are from the decay constant of
       $B_{c}$ and $B_{s}$ mesons.}
 \footnotetext[8]{It is estimated with the BSW \cite{zpc29p637} model.}
 \footnotetext[9]{It is estimated with a relativistic quark model
       based on the Bethe-Salpeter equation, where $a_{1}$ $=$ 1.2.}
 \footnotetext[10]{It is estimated with the light-cone QCD sum rules.}
 \footnotetext[11]{It is estimated with the framework of a nonrelativistic
       constituent quark model.}
 \footnotetext[12]{It is estimated with covariant quark model.}
 \footnotetext[13]{It is estimated with QCD sum rules.}
 \footnotetext[14]{It is estimated with the BSW \cite{zpc29p637} formalism,
       where uncertainty is from mass of $b$, $c$, $s$ quark and shape
       parameter ${\omega}$ of wave functions.}
 \footnotetext[15]{It is estimated by vertex sum rules.}
 \footnotetext[16]{It is estimated by overlap integral of the meson wave
       functions.}
 }
 \end{table}


 \end{document}